\documentclass[twocolumn,showpacs,aps,pra,amsmath,amssymb,superscriptaddress]{revtex4-1}
\usepackage{bm,color,bbm}
\usepackage{hyperref, mathtools,graphicx,natbib}

\newcommand{\beq}{\begin{equation}}
\newcommand{\eeq}{\end{equation}}
\newcommand{\bqa}{\begin{eqnarray}}
\newcommand{\eqa}{\end{eqnarray}}
\newcommand{\nn}{\nonumber}

\newcommand{\dg}{^\dagger}

\newcommand{\smallfrac}[2]{\mbox{$\frac{#1}{#2}$}}
\newcommand{\bra}[1]{ \langle{#1} |}
\newcommand{\ket}[1]{ |{#1} \rangle}

\newcommand{\half}{\smallfrac{1}{2}}

\newcommand{\sq}[1]{\left[ {#1} \right]}

\newcommand{\tr}[1]{{\rm Tr}\sq{ {#1} }}

\newcommand{\blk}{\color{black}}
\newcommand{\blu}{\color{blue}}
\definecolor{maroon}{rgb}{0.7,0,0}

\definecolor{ngreen}{rgb}{0.3,0.7,0.3}

\definecolor{golden}{rgb}{0.8,0.6,0.1}

\renewcommand\blu{\blk}
\begin{document}
\title{Complementarity relations for quantum coherence}
\author{Shuming Cheng}
\affiliation{Centre for Quantum Computation and Communication Technology (Australian Research Council), Centre for Quantum Dynamics, Griffith University, Brisbane, QLD 4111, Australia}
\affiliation{Key Laboratory of Systems and Control, Academy of Mathematics and Systems Science, Chinese Academy of Sciences, Beijing 100190, P. R. China}

\author{Michael J. W. Hall}
\affiliation{Centre for Quantum Computation and Communication Technology (Australian Research Council), Centre for Quantum Dynamics, Griffith University, Brisbane, QLD 4111, Australia}

%\affiliation{${}^1$Centre for Quantum Computation and Communication Technology (Australian Research Council), Centre for Quantum Dynamics, Griffith University, Brisbane, QLD 4111, Australia}
%\affiliation{${}^2$Key Laboratory of Systems and Control, Academy of Mathematics and Systems Science, Chinese Academy of Sciences, Beijing 100190, P. R. China}

%\date{\today}

% USE FOR BOTH
\begin{abstract}
Various measures have been suggested recently for quantifying the coherence of a quantum state with respect to a given basis.  We \blu first \blk use two of these, the $l_1$-norm and relative entropy measures, to investigate tradeoffs between the coherences of mutually unbiased bases.  Results include  relations between coherence, uncertainty and purity; tight general bounds restricting the coherences of mutually unbiased bases; and an exact complementarity relation for qubit coherences.  We further define the average coherence of a quantum state.  For the $l_1$-norm measure this \blu is related to a natural \blk `coherence radius' for the state, and leads to a conjecture for an $l_2$-norm measure of coherence. For relative entropy the average coherence is determined by the difference between the von Neumann entropy and the quantum subentropy of the state, and leads to upper bounds for the latter quantity. Finally, we point out that the relative entropy of coherence is a special  case of G-asymmetry, which immediately yields several operational interpretations in contexts as diverse as frame-alignment, quantum communication and metrology, and suggests generalising the property of quantum coherence to arbitrary groups of physical transformations. 
\end{abstract}

%USE FOR REVTEX
%\pacs{03.65.Ta}
\maketitle

\section{Introduction}

The notion of `coherence' in quantum mechanics is an old one, arising from the even older notion of phase coherence for classical light waves. In both cases the loss of phase correlations leads from superpositions to mixtures. For example, a projective quantum measurement in some basis will act on a pure state \blu ensemble \blk to reduce it to a mixture of \blu orthogonal \blk states, thus decreasing the coherence of the ensemble with respect to the measurement basis.  

It has recently been proposed to regard quantum coherence as a physical resource, useful for accomplishing certain tasks, that decreases under under certain physical operations such as measurement \cite{BCP14,G14,KLOJ15}. In these proposals the coherence is not a property of the quantum state alone, but is defined with respect to a given measurement basis.  Measures of coherence are required to vanish for states diagonal in this basis, corresponding to complete incoherence (in particular, they cannot be further decohered relative to this basis).  It has also been proposed they should decrease under incoherent operations, i.e.,  operations that preserve diagonality in the given basis, and to decrease on average under mixtures of such operations \cite{BCP14}.

Various suitable measures of coherence have been suggested that meet the above requirements \cite{BCP14, G14, SXFL15,YZCM15,SUBA15}.  However, while these measures are formally satisfactory, the question of whether they in fact quantify some physical resource has not been settled in most cases (an exception being the relative entropy of coherence \cite{WY15,SBMP15}; see also below).  This question has an analogy in statistical physics and communication theory: there are many possible formal measures of entropy that quantify irreversibility, but only the Gibbs and Shannon entropies (and their quantum generalisations) appear to have significance as direct physical resources.

One approach to assessing the degree to which a measure of coherence relates to a resource is to ask whether there is `only so much to go round'.  In particular, if a state has a high measure of coherence with respect to one basis, how high can its coherence be with respect to another basis? -- is a resource tradeoff involved?  

We examine this question for the case of mutually unbiased bases (MUBs), for two particular measures of coherence: the $l_1$-norm and the relative entropy \cite{BCP14}.  Here, two observables and their corresponding basis sets are said to be mutually unbiased, or `complementary', if the measurement distribution of either one is uniform for any eigenstate of the other \cite{S60, I81, W86, K87,  WF89}. We find that the $l_1$-norm measure satisfies an exact tradeoff relation for qubits, and more generally has a tight bound \blu determined by \blk the difference between a quantum and a classical purity (Sec.~II~A).  Further, the sum of the squared $l_1$-norm coherences of a maximal set of mutually unbiased bases (MUBs) has a tight  upper bound in terms of  a  `radius of coherence' for the state (Sec.~II~B). These results \blu lead to \blk the conjecture that an $l_2$-norm measure \blu of coherence may be more natural \blk from the resource point of view.  (Sec.~II~C). We also obtain  \blu a \blk nontrivial upper bound for the corresponding sum of the relative entropies of coherence, which is \blu tight for maximally-mixed states and in the limit of arbitrarily large dimensions \blk (Sec.~III).

A second question of interest is whether one can characterise the coherence of a quantum state {\it per se}, without reference to any particular basis. In this respect, for example, it has recently been shown that the minimum coherence of a \blu  {\it multipartite} \blk state, with minimisation over all possible product basis sets, is equivalent to a particular measure of discord for the state \cite{YXGS15}.  However, this result cannot be \blu extended \blk  to define the coherence of a  quantum state {\it per se}, as the minimum coherence is always zero (corresponding to a basis in which the state is diagonal).

In answer to this second question we propose \blu using an \blk average \blu measure of the \blk coherence, over all basis sets (Sec.~IV).  Such averages represent the degree to which the state is a useful coherent resource if a basis is chosen at random.  For the $l_1$-norm measure the average coherence is bounded by the coherence radius of the state. For the relative entropy of coherence the average coherence is proportional to the difference between the von Neumann entropy and the quantum subentropy \cite{DDJ14} of the state, providing a new interpretation of  the latter quantity.  We also obtain  new upper bounds for the subentropy.

Finally, in the concluding section we point out that the relative entropy of coherence is \blu related to the efficiency of quantum heat engines \cite{L97}, and is also a special case of the G-asymmetry \cite{VAWJ08}.  This yields alternative physical \blk interpretations of this quantity to those \blu proposed more recently \blk \cite{WY15, SBMP15}, and suggests generalising the property of quantum coherence to arbitrary groups of physical transformations.

\section{Complementarity for $l_1$-norm measure of coherence}

For a quantum state described by density operator $\rho$, and \blu an \blk orthonormal basis $A\equiv \{|a\rangle\}$, the $l_1$-norm measure of coherence is defined by \cite{BCP14}
\beq \label{c1}
\mathcal{C}_1(A,\rho)=\sum_{a\neq a'}|\bra{a}\rho \ket{a'}|.
\eeq
Note that the normalisation and positivity of $\rho$ yield the inequality
\begin{align} \nn
\mathcal{C}_1(A,\rho) &= \sum_{a,a'}|\bra{a}\rho \ket{a'}|-1\\ \nn
&\leq \sum_{a,a'}\bra{a}\rho \ket{a}^{1/2}\,\bra{a'}\rho \ket{a'}^{1/2} -1\\
&= \left(\sum_a \bra{a}\rho \ket{a}^{1/2}\right)^2 -1,
\end{align}
with equality for all pure states $\rho=|\psi\rangle\langle\psi|$.  For a $d$-dimensional Hilbert space it follows that the maximum possible value corresponds to a uniform probability distribution of $A$ for the state,  $\bra{a}\rho \ket{a}\equiv d^{-1}$, yielding
\beq  \label{c1max}
\mathcal{C}_1(A,\rho)\leq \mathcal{C}_1^{\rm max} := d-1.
\eeq

We now investigate the restrictions on this maximum degree of coherence, both in terms of the purity of the quantum state and  when more than one basis is considered (and for MUBs in particular).  These restrictions lead to far stronger upper bounds on individual coherences than Eq.~(\ref{c1max}), and connect coherence to uncertainty, to the difference of quantum and classical purities, and to a natural `radius of coherence'.  We first consider qubits, and then the general $d$-dimensional case.

\subsection{Identities for qubit coherences}

Let $\sigma_1$, $\sigma_2$ and $\sigma_3$ denote the three Pauli qubit observables.  These are mutually unbiased, in the sense that the distribution of any one of these observables is uniform for any eigenstate of the others \cite{S60, I81, W86, K87, WF89}. Using the same symbols for the corresponding basis sets, it is straightforward to calculate from Eq.~(\ref{c1}) and the Bloch representation $\rho=\half(1+r\cdot\sigma)$ that
\begin{align} \nn
\mathcal{C}_1(\sigma_3,\rho)^2 &= 4|\langle +|\rho|-\rangle|^2\\ \nn
&= 2\left( \sum_{z,z'= \pm 1} |\langle z|\rho| z'\rangle|^2 - \sum_{z= \pm 1} |\langle z|\rho|z\rangle|^2\right)\\ \nn
&= 2\tr{\rho^2} - 2 \sum_{z=\pm 1} \tr{\rho \frac{1+z\sigma_3}{2}}^2 \\ 
\label{qubitcoh}
& = r\cdot r - (r_3)^2.
\end{align}
One obtains similar results for $\sigma_1$ and $\sigma_2$, yielding the equality
\beq \label{c1eq}
\mathcal{C}_1(\sigma_1,\rho)^2+ \mathcal{C}_1(\sigma_2,\rho)^2+ \mathcal{C}_1(\sigma_3,\rho)^2 = 2\, r\cdot r
\eeq
for mutually unbiased qubit coherences.

The above equality is stronger than Eq.~({\ref{c1max}), and clearly constrains the usefulness of the state as a coherence resource.  For example, if the coherence is maximal with respect to $\sigma_1$ and $\sigma_2$, i.e., $\mathcal{C}_1(\sigma_1,\rho)=\mathcal{C}_1(\sigma_2,\rho)=1$, then it must vanish with respect to $\sigma_3$, i.e., $\mathcal{C}_1(\sigma_3,\rho)=0$.  
It also follows from Eq.~(\ref{c1eq}) that the qubit coherence for a given basis is constrained not only by the coherences of mutually unbiased bases, but by the length of the Bloch vector $r$.  In particular, for the maximally-mixed state with $r=0$ all coherences must vanish.

Note that while Eq.~(\ref{c1eq}) may be interpreted as a complementarity relation for qubit coherences, it should be distinguished from uncertainty relations.   In particular, noting that the mean square error of $\sigma_3$ for state $\rho$ is given by $(\Delta_\rho \sigma_z)^2=1-(r_3)^2$, Eq.~(\ref{qubitcoh}) immediately generalises to the \blu qubit relation \blk
\beq \label{certainty}
%\mathcal{C}_1(A,\rho)^2 + (\Delta_\rho A)^2= 1 + r\cdot r  %= 2\tr{\rho^2}
(\Delta_\rho A)^2 = \mathcal{C}_1(A,\rho)^2 + 1 - r \cdot r 
\eeq
\blu for \blk  coherence \blu and  uncertainty. \blk Thus, a high degree of coherence \blu implies \blk a high degree of uncertainty, and vice versa, but only for a fixed degree of purity (as defined by the length of the Bloch vector).

\subsection{General case: purity and mutually unbiased coherences}

We first generalise the qubit identity~(\ref{qubitcoh}) to arbitrary dimensions.  In particular, using the Schwarz inequality, it follows from Eq.~(\ref{c1}) that
\begin{align} \nn
\mathcal{C}_1(A,\rho)^2 &=\left(\sum_{a\neq a'}|\bra{a}\rho \ket{a'}|\right)^2 \\ \nn
&\leq d(d-1)\left(\sum_{a\neq a'}|\bra{a}\rho \ket{a'}|^2\right)      \\ \nn
&= d(d-1)\left(\sum_{a, a'}|\bra{a}\rho \ket{a'}|^2-\sum_{a}|\bra{a}\rho \ket{a}|^2\right)  \\ 
\label{ineq}
&=  d(d-1)\left(\tr{\rho ^2}-\sum_a \bra{a}\rho\ket{a}^2\right) .
\end{align}
Hence, defining the quantum and classical purities
\[ P(\rho):=\tr{\rho^2},~~~~P(A|\rho):= \sum_a \bra{a}\rho\ket{a}^2, \]
respectively, the coherence is bounded by 
\beq \label{purediff}
\mathcal{C}_1(A,\rho) \leq \sqrt{ d(d-1) \left[ P(\rho) - P(A|\rho)\right] }.
\eeq
It follows that the difference between the quantum and classical purity may be regarded as a proxy resource for coherence (see also Sec.~II~C below). This is analogous to the role played by the difference between the quantum and classical entropy for the case of the relative entropy measure of coherence (see Sec.~III).  

Equation~(\ref{purediff}) reduces to the qubit \blu identities~(\ref{qubitcoh}) and (\ref{certainty})  when \blk $d=2$.
%Further, since the maximum possible quantum purity is $P(\rho)=1$, and the minimum possible classical purity is $P(A|\rho)=d^{-1}$, it follows from Eq.~(\ref{purediff}) that $\mathcal{C}_1(A,\rho)\leq d-1$, in agreement with Eq.~(\ref{c1max}). 
More generally, noting that the classical purity is never less than $d^{-1}$, it follows immediately that 
\beq \label{singh}
\mathcal{C}_1(A,\rho)^2  \leq (d-1) \left[ d\,P(\rho) -1\right] ,
\eeq
which is equivalent to the bound in Theorem 1 of Singh {\it et al.}~\cite{SBHP15}.  Thus,  Eq.~(\ref{purediff}) is stronger than (and provides a far simpler derivation of) the latter bound.

We will now use the strong upper bound in Eq.~(\ref{purediff}) to obtain a tight `complementarity' tradeoff for the quantum coherences of a complete set of MUBs. Recall that two orthonormal basis sets $A$ and $B$, for a $d$-dimensional Hilbert space, are said to be mutually unbiased, or maximally complementary, if their overlaps are constant, i.e., if $|\langle a|b\rangle|^2=d^{-1}$ for all $a$ and $b$ \cite{S60, I81, W86,K87, WF89}. Thus, no information encoded in basis $B$ can be recovered by a measurement in basis $A$ --- the measurement completely decoheres any such encoded information. This is reflected in the property
\beq \mathcal{C}_1(A,|b\rangle\langle b|)=d-1 =  \mathcal{C}_1^{\rm max} \eeq  
for MUBs, following from definitions~(\ref{c1}) and (\ref{c1max}), i.e., eigenstates of $B$ are maximally coherent relative to $A$, and hence undergo maximum possible decoherence under a projective measurement of $A$. 

It is known  that there exists a complete set of $d+1$ MUBs, $A_1,A_2,\dots,A_{d+1}$, when the Hilbert space dimension $d$ is a prime power \cite{W86,WF89}, and we will now consider this case (in Sec.~IV we will give a related result holding for arbitrary dimensions $d$).
 Ivanovic showed that a such complete set is useful for quantum state tomography: one has the beautiful identity \cite{I81}
\beq \label{tom}
\rho =  \sum_j \rho(A_j) -\hat 1,
\eeq
where the density operator $\rho(A)$ is defined by
\[ \rho(A) := \sum_a |a\rangle\langle a|\, \langle a|\rho|a\rangle \]
for basis set $A$ and state $\rho$. 
For qubits, Eq.~(\ref{tom}) corresponds to reconstructing the components of the Bloch vector from measurements of $\sigma_1,\sigma_2$ and $\sigma_3$, \blu while \blk more generally the measurement distributions of $A_1,\dots,A_{d+1}$ suffice for reconstruction of the state.  From \blu the above \blk identity one can easily derive the relation \cite{I92}
\beq \label{pureid}
\sum_{j=1}^{d+1} P(A_j|\rho) = 1 + P(\rho),
\eeq
\blu connecting the individual \blk  classical purities to the quantum purity.  

\blu We now sum over the square of Eq.~(\ref{purediff}) for a set of MUBs $A_1,\dots,A_{d+1}$, and substitute Eq.~(\ref{pureid}) into the result,  to obtain \blk the complementarity relation
\beq \label{comp}
\sum^{d+1}_{j=1}\mathcal{C}_1(A_j,\rho)^2 \leq d(d-1) \,\left[d P(\rho) -1\right].
\eeq
for mutually unbiased coherences.  \blu This \blk relation is stronger than that obtained by summing over the weaker bound in Eq.~(\ref{singh}), \blu and implies, in particular, \blk that at most $d$ of the $d+1$ MUBs can simultaneously achieve the maximal possible coherence $\mathcal{C}_1^{\rm max} =d-1$ in Eq.~(\ref{c1max}), with the remaining coherence forced to vanish. 

The complementarity relation~(\ref{comp}) is in fact tight, in the sense that it is saturated by some quantum state for any given value of the quantum purity. For example, for pure states, \blu with maximum purity \blk $P(\rho)=1$, the bound reaches its maximum possible value of $d(d-1)^2$ and is saturated by choosing $\rho$ to \blu to be any one \blk of the basis states. Conversely, for the maximally mixed state $\rho=d^{-1}\hat 1$, \blu with minimum purity \blk $P(\rho)=d^{-1}$, both sides of the relation vanish.
More generally, Eq.~(\ref{comp}) is saturated by the states of the form
\beq \label{example}
\rho_\epsilon = (1-\epsilon) |b\rangle\langle b| + \frac{\epsilon}{d-1}\left(\hat 1 - |b\rangle\langle b|\right),
\eeq
where $|b\rangle$ is an element of any of the $d+1$ MUBs and $0\leq\epsilon\leq 1$. \blu These \blk states vary continuously from the pure state $|b\rangle\langle b|$ to the maximally-mixed state $d^{-1}\hat 1$, for $\epsilon\in[0,1-d^{-1}]$, and hence achieve all possible values of the quantum purity.  

\blu The \blk saturation of Eq.~(\ref{comp}) \blu by the states  $\rho_\epsilon$ \blk follows directly from the saturation of the Schwarz inequality in the second line of Eq.~(\ref{ineq})  ---  the only point at which an inequality enters the derivation of the \blu complementarity relation \blk  --- by any state satisfying
$|\langle a|\rho|a'\rangle|=$constant, for all  $a\neq a'$.
In particular, if $A=A_j$ is the basis set that contains $|b\rangle$ then $b=a_0$ for some $a_0$ and the off-diagonal elements all vanish; otherwise $|b\rangle$ must be from a  basis set mutually unbiased with respect to $A$ and so $|\langle a|\rho|a'\rangle|=\epsilon[1-(d-1)^{-1}]|\langle a|b\rangle\langle b|a'\rangle|=\epsilon (d-1)(1-2d^{-1})$ \blu which \blk is again constant for $a\neq a'$. This argument can also be used to show that the bound in Eq.~(\ref{singh}) is saturated by $\rho_\epsilon$, with $|b\rangle$ chosen from any basis set mutually unbiased to $A$, greatly simplifying the derivation \blu of Theorem 2 \blk in \cite{SBHP15}.

\subsection{Radius of coherence and a conjecture}

It is of interest to note that the complementarity relation for mutually unbiased coherences in Eq.~(\ref{comp}) can be rewritten in the geometric form
\beq \label{geom}
\sum^{d+1}_{j=1}\mathcal{C}_1(A_j,\rho)^2 \leq R_1(\rho)^2,
\eeq
with
\beq \label{radius}
R_1(\rho):= \sqrt{d(d-1) \,\left[d P(\rho) -1\right]}.
\eeq
Thus, the coherences $\mathcal{C}_1(A_j,\rho)$ are constrained to lie on or within a hypersphere of radius $R_1(\rho)$.  We will call this the {\it radius of coherence} of the state. It is maximal for pure states, and vanishes for the maximally-mixed state.  

The radius of coherence may be thought of as quantifying the `intrinsic' coherence of the state as a resource, independently of any particular basis set.  
This resource places a strict bound on coherences of a complete set of MUBs via Eq.~(\ref{geom}). More generally, as will be shown in Sec.~IV, it bounds the average coherence over all basis sets, for any Hilbert space dimension $d$, whether or not a complete set of MUBs exists.  

Equations (\ref{purediff}) and (\ref{geom}) further suggest that the quantity
\beq
\mathcal{C}_2(A_j,\rho) := \left( \sum_{a\neq a'}|\bra{a}\rho \ket{a'}|^2\right)^{1/2}
\eeq
is a very natural candidate for a measure of coherence.  We will call this quantity, for obvious reasons, the \blu  $l_2$-norm \blk measure.  The square of this quantity has been previously considered as a possible coherence measure, but rejected as it was shown by a counterexample to not satisfy all of the required properties mentioned in the Introduction \cite{BCP14}.  However, this counterexample fails for $\mathcal{C}_2(A_j,\rho)$ itself, and we conjecture that this quantity does satisfy  the necessary properties. It is easy to check that the $l_2$-norm measure vanishes if and only if only $\rho$ is diagonal with respect to the basis $A$, and that it is convex with respect to $\rho$ (since it is the matrix norm of the difference between $\rho$ and its diagonal in the $A$ basis). \blu However, \blk
%\cheng{$\mathcal{C}_2$ inherently possesses the properties: 1)~$\mathcal{C}_2(A,\rho)=0$ if and only if $\rho$ is diagonal in basis of $A$; 2)~2-norm is a good metric, so $\mathcal{C}_2(A,\rho)$ is convex. Although counterexamples are given to prove that the distance based on squared 2-norm is not a valid entanglement measure \cite{O00} and not a well-defined discord measure \cite{P12} and contractivity does not hold in the case of qutrits for 2-norm under the positive trace preserving maps \cite{PWPM06}, further
further efforts are needed to \blu determine \blk whether or not the remaining requirements proposed in \cite{BCP14} (relating to the decrease of coherence measures under incoherent operations), are also satisfied. 

The main advantage of the $l_2$-norm measure, from the point of view of coherence as a resource, is that all of the inequalities derived \blu above \blk for $\mathcal{C}_1(A_j,\rho)$ become strict equalities for $\mathcal{C}_2(A_j,\rho)$.  In particular, the derivations of Eqs.~(\ref{purediff}) and (\ref{comp})  \blu lead directly \blk to 
\beq \label{c2}
\mathcal{C}_2(A,\rho) = \sqrt{ \left[ P(\rho) - P(A|\rho)\right] },
\eeq
\beq \label{com2}
\sum^{d+1}_{j=1}\mathcal{C}_2(A_j,\rho)^2 = d P(\rho) -1 =: R_2(\rho)^2,
\eeq
where $ R_2(\rho)$ is a coherence radius analogous to (and proportional to) $R_1(\rho)$ in Eq.~(\ref{geom}).  Hence, if the conjecture is valid, the difference of the quantum and classical purity moves from being merely a proxy coherence measure in Eq.~(\ref{purediff}) to (in square root form) a genuine coherence measure in Eq.~(\ref{c2}).  Further, the complementarity of the coherences of a complete set of MUBs becomes precisely captured, by the geometric property that they must lie on a hypersphere of radius $R_2(\rho)$ \blu as per Eq.~(\ref{com2}). \blk

\section{Complementarity for relative entropy measure of coherence}

For a quantum state described by density operator $\rho$ and an orthonormal basis $A\equiv \{|a\rangle\}$, the relative entropy measure of coherence is defined by \cite{BCP14}
\begin{align} \label{crel}
\mathcal{C}_{\rm rel}(A, \rho)& :=H(A|\rho) - S(\rho) \leq \log d - S(\rho),
\end{align} 
where $H(A|\rho):= -\sum_a\bra{a}\rho\ket{a}\log\bra{a}\rho\ket{a}$ is the Shannon entropy of the probability distribution of $A$ for state $\rho$, and $S(\rho):=-\tr{\rho \log \rho}$ is the von Neumann entropy of $\rho$.  Note that the base of the logarithm in Eq.~(\ref{crel}) is arbitrary, corresponding to a choice of units, with base 2 corresponding to units of bits. \blu It is seen that \blk $\mathcal{C}_{\rm rel}(\rho)$ is the difference between a quantum and a classical entropy, providing an interesting analogy to the difference of quantum and classical purities in Eqs.~(\ref{purediff}) and (\ref{c2}).

To obtain a complementarity relation for the coherences of a complete set of $d+1$ MUBs, we will make use of the entropic certainty relation  \cite{S95} 
\begin{align} \nn
\sum^{d+1}_{j=1} H(A_j|\rho) \leq & (d+1) \log d\\
\label{srd}~& - \frac{(d-1)\left[ d\,P(\rho)-1 \right]}{d(d-2)} \log (d-1),
\end{align}
For qubits the upper bound reduces to $3\log 2 - \left[ P(\rho)-\frac{1}{2} \right]\log e$ by taking the continuous limit $d\to 2$.  This can be improved (by up to $\approx 2\%$), to  the tight qubit relation \cite{S95}
\beq \label{sr2}
\blu \sum_{j=1}^3 H(\sigma_j|\rho) \blk \leq h\left( \sqrt{[2P(\rho)-1]/3}\right) ,
\eeq
for the MUBs corresponding to the Pauli spin matrices $\sigma_1,\sigma_2,\sigma_3$, \blk with \blk $h(x):=-\frac{1+x}{2}\log \frac{1+x}{2} -\frac{1-x}{2}\log \frac{1-x}{2} $.

Equations (\ref{crel}) and (\ref{srd})  immediately yield the complementarity relation
 \begin{align} \nn
 \sum^{d+1}_{j=1}\mathcal{C}_{\rm rel}(A_j,\rho) \leq &
 %&\leq(d+1)\left[\log d-S(\rho)\right]-\frac{\tau_d}{d-1}\sum^{d+1}_{j=1} \,\left[ d\,P(A_j|\rho)-1 \right]\log d \\ 
 %&=(d+1)[\log d-S(\rho)]-\frac{\tau_d}{d-1} \,\left[ d\,P(\rho)-1 \right]\log d  %\\ 
(d+1) \left[\log d -S(\rho)\right] \\
~& \blu -  \frac{(d-1)\left[ d\,P(\rho)-1 \right]}{d(d-2)} \log (d-1) \blk
\label{ent1}
% &=(d+1)[\log d-S(\rho)]-\frac{\tau_d}{d-1} \,R_2(\rho)^2\log d
 \end{align}
for the coherences of a complete set of MUBs.  The first term in the upper bound corresponds to summation over the trivial bound in Eq.~(\ref{crel}).  Hence, the subtraction of the second term generates a nontrivial bound for the sum of the coherences.  
 This bound is tight for the maximally-mixed state, with both sides of the inequality vanishing.  It is also tight for pure states in the limit $d\to\infty$.  In particular, in this limit the upper bound approaches $d\log d$, which is saturated by choosing $\rho$ to correspond to any one of the basis elements in $A_1,\dots A_{d+1}$.  More generally, the bound becomes more closely achievable as $d$ increases, for any given value of the purity. 

For qubits, Eqs.~(\ref{crel}) and (\ref{sr2}) yield the stronger complementarity relation
\beq \label{ent2} 
\sum_{j=1}^3\mathcal{C}_{\rm rel}(\sigma_j,\rho)\leq 3 \left[h\left( \sqrt{[2P(\rho)-1]/3}\right)-S(\rho)\right] .
\eeq
This relation is tight in the sense that it is saturated by some state $\rho$ for any given value of the quantum purity $P(\rho)$ (in particular, one may choose $\rho$ to have equal Bloch vector components $r_j=\sqrt{(2P(\rho)-1)/3}$, which saturates the certainty relation in Eq.~(\ref{sr2}) \cite{S95}.

Finally, we note that the coherence radiuses $R_1(\rho)$ and $R_2(\rho)$ from Sec.~III~C appear naturally in the complementarity relations~(\ref{ent1}) and (\ref{ent2}).

\section{Average quantum coherences} 

As mentioned in the Introduction, it is of interest to characterise the coherence of a quantum state {\it per se}, without reference to any specific basis. We propose using an average of the coherence over all basis sets. Such an average may be interpreted as \blk the degree to which the state is a useful coherent resource for a randomly chosen measurement basis. There are, however, many ways to define averages.  We will consider both the mean coherence and the root mean square coherence, defined for any given measure of coherence $\mathcal{C}(A,\rho)$ by
\beq \label{av}
 \overline{\mathcal{C}}(\rho):=\int dU\mathcal{C}(UAU^\dag,\rho) ,
 \eeq
\beq \label{ms}
{\rm RMS}[\mathcal{C}(\rho)] := \left[ \int dU\, \mathcal{C}(UAU^\dag,\rho)^2 \right]^{1/2},
\eeq
respectively. Here $U$ ranges over the group of unitary transformations (where any two basis sets are connected by such a transformation), and  $dU$ denotes the normalised invariant Haar measure over this group \cite{S74}.  
\blu Note that the  \blk convexity of the function $f(x)=x^2$ implies the relation
\beq \label{var}
\overline{\mathcal{C}}(\rho) \leq {\rm RMS}[\mathcal{C}(\rho)] .
\eeq
Hence, any upper bound for the root mean square coherence is also a bound for the mean coherence.

We  will first investigate average coherences for  the $l_1$-norm and $l_2$-norm measures of coherence considered in Sec.~II, and then for the relative entropy measure of coherence considered in Sec.~III. Bounds for these average coherences are closely related to the complementarity relations obtained in previous sections, \blu and in particular to the coherence radiuses $R_1(\rho)$ and $R_2(\rho)$. \blk  However, the \blu bounds we obtain \blk have the advantage of being applicable to all Hilbert space dimensions $d$, whereas complete MUBs are only known to exist for the case that $d$ is a prime power.

 \subsection{Average $l_1$-norm and $l_2$-norm measures of coherence}
 
Averages of the $l_1$-norm and $l_2$-norm measures cannot be evaluated analytically, with the exception of the root mean square average for the $l_2$-norm as per below.  However, upper bounds may be obtained, using the results of previous sections together with the identity
\beq \label{intid}
\int dU\,{P(UAU^\dag|\rho)} = \frac{1+P(\rho)}{1+d}
\eeq
following from Eq.~(10) of Ref.~\cite{H00}.  Note that, comparing with Eq.~(\ref{pureid}), this identity shows that the average classical purity over all basis sets is equal to the average over a complete set of MUBs (whenever such a set exists).

Now, from Eqs.~(\ref{purediff}) and (\ref{ms}) it follows that
\begin{align*}
{\rm RMS}[\mathcal{C}_1(\rho)]^2 
&\leq d(d-1) \left[ P(\rho) - \int dU P(UAU^\dag|\rho)\right], 
\end{align*}
while from Eqs.~(\ref{c2}) and (\ref{ms}) one has the identity
\beq \nn
 {\rm RMS}[\mathcal{C}_2(\rho)]^2 \blu = \blk  P(\rho) - \int dU P(UAU^\dag|\rho). 
\eeq
Using Eqs.~(\ref{var}) and (\ref{intid}) then  yields the upper bounds
\begin{align}
\label{c1av}
\overline{\mathcal{C}}_1(\rho) &\leq{\rm RMS}[\mathcal{C}_1(\rho)]\leq \frac{R_1(\rho)}{\sqrt{d+1}} ,\\
\label{c2av}
\overline{\mathcal{C}}_2(\rho) &\leq {\rm RMS}[\mathcal{C}_2(\rho)] = \frac{R_2(\rho)}{\sqrt{d+1}} 
\end{align}
for the average $l_1$-norm and $l_2$-norm coherences, in terms of the radiuses of coherence defined in Eqs.~\ref{radius}) and (\ref{com2}).   

Equations~(\ref{c1av}) and (\ref{c2av}) hold whether or not a complete set of MUBs exists, and so may be regarded as generalisations of Eqs.~(\ref{radius}) and (\ref{com2}) to arbitrary dimensions.
Further, the result for the root mean square coherence of the $l_2$-norm in Eq.~(\ref{c2av}) is an equality rather than an upper bound, and is directly proportional to the corresponding radius of coherence. This result provides additional motivation for the conjecture in Sec.~III~C, that $\mathcal{C}_2(A,\rho)$ satisfies all of the requirements for a measure of coherence.

\subsection{Average relative entropy measure of coherence}

The mean relative entropy measure of coherence follows from Eqs.~(\ref{crel}) and (\ref{av}) as
\beq
\overline{\mathcal{C}}_{\rm rel}(\rho)=\int dU\,H(UAU^{\dg}|\rho) - S(\rho).
\eeq
The mean entropy over all basis sets may be expressed in terms of the quantum subentropy $Q(\rho)$ by \cite{W90,JRW94,DDJ14}:
\beq \label{entav}
\int dU\,H(UAU^{\dag}|\rho) = Q(\rho) + C_d,
\eeq
where  $C_d:=(\frac{1}{2} +\frac{1}{3}+ \dots +\frac{1}{d})\log e$, and
\beq \label{qrho}
Q(\rho) := -\sum_{i=1}^d \left(\prod_{i\neq j}\frac{\lambda_i}{\lambda_i-\lambda_j}\right)\lambda_i \log \lambda_i,
\eeq
in terms of the eigenvalues $\{\lambda_1,\dots,\lambda_d\}$ of $\rho$. 

The quantum subentropy is a tight lower bound on the accessible information of pure-state ensembles, and is never greater than the von Neumann entropy $S(\rho)$ \cite{JRW94,DDJ14}.   It follows from the above that
\beq \label{subby}
\overline{\mathcal{C}}_{\rm rel}(\rho) = C_d - \left[S(\rho) - Q(\rho) \right]  ,
\eeq
providing an alternative interpretation of the quantum subentropy in terms of quantum coherence. In particular, the coherence of state $\rho$, as quantified by $\overline{\mathcal{C}}_{\rm rel}(\rho)$, is determined by the difference between the von Neumann entropy and the subentropy.   A maximum coherence of $C_d$ is obtained for pure states, for which $S(\rho)=Q(\rho)=0$, while a mininum coherence of 0 is obtained for the maximally-mixed state.

The subentropy is a rather complicated function of the eigenvalues of the quantum state, and is nontrivial to evaluate when $\rho$ has degenerate eigenvalues \cite{W90,JRW94,DDJ14} (see also below).  Hence, it is of interest to bound the average coherence in Eq.~(\ref{subby}) via a corresponding bound on the subentropy. For example, the known bound $Q(\rho)\leq \log d-C_d$ \cite{JRW94} yields $\overline{\mathcal{C}}_{\rm rel}(\rho)\leq \log d - S(\rho)$.  However, this not particularly strong, and indeed follows immediately by taking the average of the inequality in Eq.~(\ref{crel}).  A stronger upper bound for  the average coherence and the subentropy may be obtained using either the entropic certainty relation (\ref{srd}) or the complementarity relation (\ref{ent1}), whenever a complete set of MUBs exists.  For example, averaging the former relation over density operators $U^\dag\rho U$ with respect to the Haar measure, and noting $H(A|U^\dag\rho U)=H(UAU^\dag|\rho)$, yields
\begin{align*} \nn
\sum_{k=1}^{d+1} \int dU\,H(UA_kU^{\dag}|\rho) &\leq (d+1) \log d\\
&- \frac{d-1}{d(d-2)} \left[ d\,P(\rho)-1 \right] \log (d-1) .
\end{align*}
The left hand side is just $d+1$ times the left hand side of Eq.~(\ref{entav}), yielding the bound
\beq \label{qupper}
Q(\rho) \leq \log d - C_d -\frac{(d-1)\left[ d\,P(\rho)-1 \right]}{d(d+1)(d-2)} \log (d-1) 
\eeq
for the quantum subentropy, and a corresponding upper bound for the average coherence via Eq.~(\ref{subby}). By inspection, this is stronger than the  bound $Q(\rho)\leq \log d - C_d$ \cite{JRW94} (with equality only for the maximally-mixed state).  It is also stronger than the recent bound $Q(\rho) \leq -\log \lambda_{\rm max}(\rho)$ derived in Ref.~\cite{DDJ14}, for sufficiently mixed states (where $\lambda_{\rm max}(\rho)$ denotes the maximum eigenvalue of $\rho$), as illustrated in Fig.~1 for the cases $d=2$ and $d=11$.  A marginally stronger bound for $d=2$ may be similarly obtained, using Eq.~(\ref{sr2}).

The upper bound in Eq.~(\ref{qupper}) is valid when there is a complete set of MUBs, where such sets are only known to exist when $d$ is a prime power \cite{W86,WF89}.  It is plausible that the bound in fact holds for all dimensions $d$.  However, it is possible to obtain a weaker bound that is certainly valid for all dimensions, based on a result by Harremo\"es and Tops\o{}e relating classical entropies and purities. In particular, from Theorem II.8 and Corollary II.9 of Ref.~\cite{HT01} one has
\beq \label{harr}
H(A|\rho) \leq \left\{ 1 -\frac{\tau_d}{d-1} \,\left[ d\,P(A|\rho)-1 \right]\right\} \log d,
\eeq
where $\tau_d$ is a strictly increasing sequence with $\tau_2 =(\ln 4)^{-1}\approx 0.7213$ and $\lim_{d\to\infty}\tau_d = 1$. Lemma VI.8 of \cite{HT01} further gives the analytic lower bound  $\tau_d\geq 1- (1+ \ln d)^{-1}$.  Replacing $A$ by $UAU^\dag$ in this inequality, integrating over $U$ with respect to the Haar measure, and using Eqs.~(\ref{intid}) and (\ref{entav}), yields the general result
\beq \label{ht}
Q(\rho) \leq \log d - C_d - \frac{\tau_d\left[dP(\rho)-1\right]}{d^2-1} \log d 
\eeq
valid for all dimensions.  A corresponding upper bound follows for the average coherence via Eq.~(\ref{subby}). Note that $\tau_d$ may be replaced by its upper bound $1- (1+ \ln d)^{-1}$ for more easily evaluable bounds.  

\begin{figure}
\centering
\includegraphics[width=0.45\textwidth]{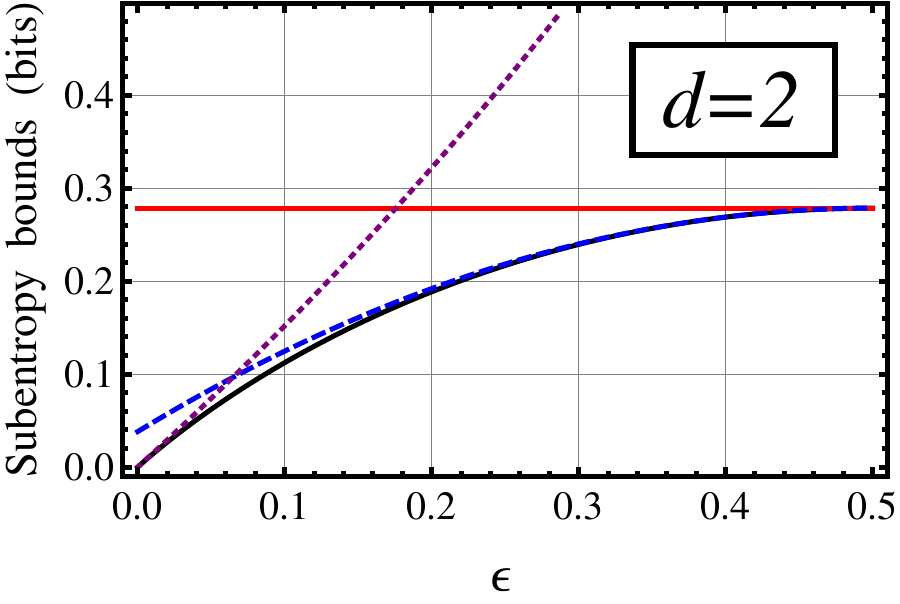}\\
~\\
\includegraphics[width=0.45\textwidth]{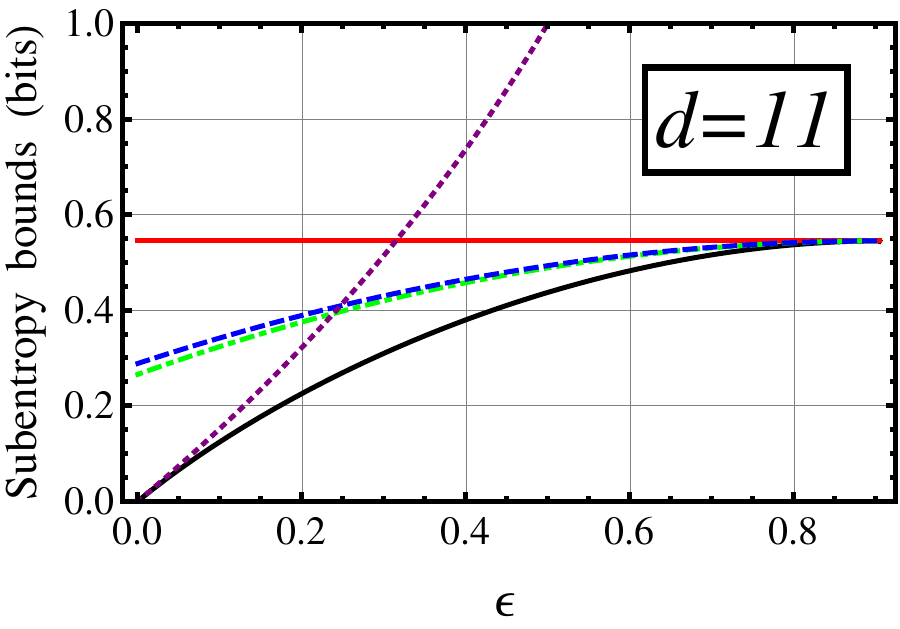}
\caption{Bounds for subentropy for $d=2$ and $d=11$, plotted for the states $\rho_\epsilon$ in Eq.~(\ref{example}) for $\epsilon\in[0,1-1/d]$.  Note that $\epsilon=0$ corresponds to a pure state, and $\epsilon=1-1/d$ to a maximally-mixed state. The lower black curve in each subfigure shows the exact value of the subentropy, $Q(\rho_\epsilon)$, in Eq.~(\ref{tricky}); the green dash-dotted and blue dashed curves show the upper bounds in Eqs.~(\ref{qupper}) and (\ref{ht}) respectively (identical for the $d=2$ case); the horizontal red curve is the known upper bound $\log d-C_d$ in Ref.~\cite{JRW94}, and the dotted purple curve is the known upper bound $-\log \lambda_{\rm max}$ in Ref.~\cite{DDJ14}. The bounds may also be used to bound the average relative entropy of coherence in Eq.~(\ref{subby}), as discussed in the main text. }
\end{figure}

The performance of the bounds (\ref{qupper}) and (\ref{ht})  for subentropy is exhibited in Fig.~1, using the states $\rho_\epsilon$ in Eq.~(\ref{example}), where these states range from a pure state to the maximally-mixed state.  The bounds are seen to be stronger than the known bound $Q(\rho)\leq \log d - C_d$ \cite{JRW94} (the horizontal line in the Figure), and also significantly stronger than the known bound $Q(\rho) \leq -\log \lambda_{\rm max}(\rho)$ \cite{DDJ14} for sufficiently mixed states.  Corresponding bounds for $\overline{\mathcal{C}}_{\rm rel}(\rho)$ immediately follow via Eq.~(\ref{subby}). 

Finally, we note that the usefulness of such bounds is emphasised by the fact that the calculation of $Q(\rho_\epsilon)$ in Fig.~1 is highly nontrivial.  In particular, since $d-1$ of the eigenvalues of $\rho_\epsilon$ are degenerate, it is necessary to use the contour integral representation of $Q(\rho)$ \cite{JRW94},
	\beq
	Q(\rho)=\frac{1}{2\pi i}\oint\frac{z^d \log_2 z~dz}{{\rm det}\left(I-\rho/z\right)},
	\eeq
with the contour containing the non-zero eigenvalues of $\rho$, to calculate
\beq \label{tricky}
Q(\rho_\epsilon) = -\frac{\lambda^d_1~\log \lambda_1}{(\lambda_1-\lambda_2)^{d-1}}-\frac{1}{(d-2)!}\left(\frac{d}{d\lambda_2}\right)^{d-2}\,\frac{\lambda_2^d\log \lambda_2}{\lambda_2-\lambda_1}
\eeq
with $\lambda_1=1-\epsilon$, $\lambda_2=\epsilon/(d-1)$, which is not readily evaluable for large $d$. %\michael{Shuming: please put in the explicit formula for $Q(\rho_\epsilon)$ as the sum of two terms, where the second involves an explicit $(d-2)$-order derivative.}

\section{Conclusions}
 
We have obtained relations between uncertainty, purity and coherence, and have shown that the $l_1$-norm and relative entropy measures of coherence quantify resources in the sense of satisfying the complementarity relations and identities in Secs.~II and III for complete sets of MUBs.  In particular, the corresponding coherences cannot be simultaneously maximised.  We have also shown that the coherence radiuses $R_1(\rho)$ and $R_2(\rho)$ defined in Sec.~III~C are natural measures of the coherence of a quantum state {\it per se}, that determine tight upper bounds for MUB coherences as well as upper bounds for average coherences. These bounds reduce to identities for the $l_2$-norm measure defined in Eq.~(\ref{c2}), leading to the conjecture that this quantity, the square root of the difference between a quantum purity and a classical purity, satisfies the necessary requirements for coherence measures \cite{BCP14}.  Finally, we have shown that the average relative entropy of coherence is determined by the difference between the von Neumann entropy and the quantum subentropy, and have obtained nontrivial upper bounds for the latter quantity.

We conclude by drawing attention to previous contexts in which the relative entropy of coherence in Eq.~(\ref{crel}) has appeared, where these contexts provide futher interpretations of this quantity as a resource in addition to the operational interpretations proposed recently \cite{WY15, SBMP15}. First, noting that  $\mathcal{C}_{\rm rel}(A,\rho)$ is the entropy increase due to a measurement in basis $A$ on state $\rho$, Lloyd has shown that  measurements and similar decoherence processes  reduce the Carnot efficiency of quantum heat engines by an amount proportional to the relative entropy of coherence \cite{L97}.  Second, the relative entropy of coherence is a special case of the asymmetry of a quantum state with respect to a given group of operations $G$ --- the so-called $G$-asymmetry \cite{VAWJ08}. In particular, $\mathcal{C}_{\rm rel}(A,\rho)$ is equal to the $G$-asymmetry of $\rho$ under the group of unitary transformations that are diagonal with respect to $A$, implying that the concepts of coherence and asymmetry are equivalent in this case. For example, it is known that the $G$-asymmetry  is equal to the Holevo bound on accessible information, for a quantum communication channel corresponding to equally-weighted signal states generated by applying elements of $G$ to state $\rho$ \cite{V06, GMS09} (where this bound is achievable in the limit of arbitrarily long signals); and that it quantitatively characterises the ability of a quantum system to act as a reference frame \cite{VAWJ08,GMS09,SG12}, and as a probe state in quantum metrology \cite{HW12}. Thus, the relative entropy of coherence immediately inherits corresponding resource interpretations from these contexts.  

It is of some interest to proceed in the opposite direction, and use the notion of $G$-asymmetry to generalise the concept of quantum coherence, from basis sets to groups of physical transformations (and beyond).  In particular, for such a group $G$, let $S_G$ denote the set of states that are invariant under the elements of $G$.  Further, let $M_G$ denote the set of measurement operations with post-measurement states invariant under $G$, i.e.  such that the post-measurement ensemble $\{p_m,\rho_m\}$ for any state $\rho$ satisfies $\rho_m\in S_G$.  One can then interpret states in $S_G$ and measurement operations in $M_G$ as `incoherent' with respect to $G$.  

This interpretation naturally leads one to define a function $\mathcal{C}_G(\rho)$ to be a measure of the coherence with respect to $G$, or a `$G$-coherence', if and only if (i) coherence vanishes for incoherent states, i.e., $\mathcal{C}_G(\rho)=0$ for all $\rho\in S_G$;  (ii) coherence decreases on average under incoherent measurement operations,  i.e., $\sum p_m\,\mathcal{C}_G(\rho_m) \leq \mathcal{C}_G(\rho)$ for all post-measurement ensembles  generated by elements of $M_G$; and (iii) coherence decreases under mixing, i.e., $\mathcal{C}_G(\sum_j p_j\rho_j) \leq \sum_j p_j\,\mathcal{C}_G(\rho_j)$ for  arbitrary ensembles $\{p_j,\rho_j\}$. The second property distinguishes coherence from the broader notion of asymmetry \cite{VAWJ08, GMS09}.  These properties generalise those for $\mathcal{C}(A,\rho)$ in Ref.~\cite{BCP14}, where $G$ corresponds in this case to the group of unitary transformations diagonal with respect to the basis $A$.  

For example, the minimum quantum relative entropy, $\min_{\sigma\in S_G}S(\rho\|\sigma)$, satisfies the above properties, and is hence a suitable $G$-coherence measure. For compact unitary groups it reduces to the $G$-asymmetry $S(\rho\|\rho_G)$ (with $\rho_G:=\int_G d\mu_G\,g(\rho)$) where $d\mu_G$ is the Haar measure for $G$) \cite{VAWJ08,GMS09}.  Possible future work includes investigating groups of interest in quantum information and computation such as angular momentum and stabiliser groups, and going beyond groups to consider coherence (and/or asymmmetry) relative to an arbitrarily-defined set $S$ of incoherent states.

%\michael{V06: J. A. Vaccaro, in Proceedings of the 8th International
%Conference on Quantum Communication, Measurement
%and Computing, edited by O. Hirota, J. H. Shapiro, and
%M. Sasaki (NICT, Tokyo, 2006), pp. 421–424 (also at arXiv:1012.3532v1);  GMS09: Phys. Rev. A 80, 012307 (2009); SG12: New Journal of Physics, volume 14, 073022, 2012}

\acknowledgments
We thank Howard Wiseman for helpful discussions.  This work was supported by the ARC Centre of Excellence CE110001027. S.C. acknowledges additional support from the National Natural Science Foundation (NNSF) of China, under Grants 61134008 and 61227902.

%\appendix
%
\bibliographystyle{apsrev4-1}
\bibliography{re}

\end{document}